\documentclass[12pt]{article}
\usepackage{authblk}
\usepackage[dvips]{graphicx}
\usepackage{subfigure}
\usepackage{amsmath}

\title{Twisted space-time reduced model of large N QCD with two adjoint Wilson fermions}

\author[a,b]{Antonio Gonz\'alez-Arroyo}
\author[c]{Masanori Okawa}
\affil[a]{Instituto de F\'{\i}sica Te\'orica UAM/CSIC}
\affil[b]{Departamento de F\'{\i}sica Te\'orica, C-15 \\
              Universidad Aut\'onoma de Madrid, E-28049--Madrid, Spain}
\affil[c]{Graduate School of Science, Hiroshima University\\
Higashi-Hiroshima, Hiroshima 739-8526, Japan}

\begin{document}

\maketitle

\begin{abstract}
The space-time reduced model of large N QCD with two adjoint Wilson fermions is 
constructed by applying the symmetric twist boundary conditions with 
non-vanishing flux $k$. For large but finite  $N=L^2$, the model should
behave as the large N version of the ordinary lattice gauge model on a $V=L^4$ 
space-time volume. We perform a comparison of
the  $N$ dependence of several quantities in this model and in the 
$k$=0 model~(corresponding to periodic boundary conditions). 
Although the Z$^4$(N) symmetry seems unbroken in all cases, the N-dependence 
analysis favours the use of the same values of $k$ and $L$ for 
which the symmetry is also unbroken in the pure gauge case. In
particular, the $k=0$ model, studied recently by several
authors, shows a large and irregular dependence on $N$ within our
region of parameters. This makes this reduced model very  impractical
for  extracting physical  information about the large N lattice theory. 
On the contrary, the model for  $N=289=17^2$ and large enough $k$
gives consistent  results, even for extended observables as Wilson
loops $W(R,T)$ up to $R,T$=8, matching the expected behaviour for the
lattice model  with a   $17^4$ space-time volume.
\end{abstract}

\vspace*{-23cm}
\hspace*{8.85cm} HUPD-1303\\
\hspace*{9.5cm} IFT-UAM/CSIC-13-050\\
\hspace*{9.5cm} FTUAM-13-8

%\maketitle

\newpage

Space-time reduction is one of the most remarkable properties of  large N QCD.
Eguchi and Kawai (EK) showed\cite{EK} that the Schwinger Dyson equations obeyed by
Wilson loops in the large N lattice gauge theory are equivalent to those 
of the theory defined on one-site, provided that the global Z$^4$(N) symmetry of 
the one-site model is not broken.  Soon after, it was realized 
that the Z$^4$(N) symmetry is spontaneously broken in the weak coupling region\cite{QEK}.
To cure this problem, two modified models were proposed.  One is the quenched EK
model\cite{QEK} which, however, has recently been  shown not to work due to the hidden correlation 
between quenched phases~\cite{BS}.  The other model is the twisted
Eguchi-Kawai (TEK) model\cite{TEK}  proposed by the present authors.  The original EK model can be thought 
as the lattice theory defined on one site with  periodic boundary
conditions.  On the other hand, the TEK model  corresponds to a
one-site model with `t Hooft twisted boundary conditions. Choosing the
twist tensor appropriately, one can ensure that the ground state of the
model respects enough symmetry to guarantee reduction to work at weak
coupling. A particularly attractive choice is to take  $N$ to be the
square of an integer ($N=L^2$), and  a twist tensor that maintains the 
symmetry among the 4 space-time directions.  This is called {\em
symmetric twist} and depends on an integer $k$, which characterizes the flux
through each plane. In Ref.~\cite{TEK} we restricted our numerical
results to the simplest non-zero value $k=1$. A few  years ago, 
several authors reported signals of symmetry breaking for this version
of the model at the intermediate coupling region and for $N>100$
\cite{IO, TV, AHHI}.
The interpretation for this phenomenon is that the higher entropy of
the  Z$^4$(N)-breaking vacuum configuration overtakes the  energy gap with the
Z$^4$(N)-symmetric ground state, making the former dominate  the path integral. In a
recent paper~\cite{TEK2} we observed that the problem can be solved by
tuning $k$ to be proportional to  $L$, as the latter  is taken to infinity. This 
makes the energy gap proportional to $N^2$ and can even turn the
non-symmetric vacuum unstable. Indeed, all the results obtained with
this prescription show no signs of symmetry breaking, despite having
reached values of $N$ which are one order of magnitude larger than
before. Furthermore, we have recently performed a detailed analysis,
based on  Wilson  loops, of the reduced model and of the gauge theory  
model on a $32^4$ lattice  and several values of $N$. Our results show an 
spectacular agreement for several observables, including the string
tension, between the values obtained  from  the reduced model and  
the infinite $N$ extrapolation of the lattice gauge theory values~\cite{TEK3, TEK4}.

One of the important properties of the twisted reduced model is that it 
gives information about the size and physical interpretation of the 
finite $N$ corrections. Studying the $N=L^2$ model in perturbation
theory, one obtains that the gluon propagator becomes identical to that of the
ordinary lattice gauge  theory formulated on a finite lattice with
volume $V=L^4$~\cite{TEK}. This is the main N-dependence but not the
only one. In addition, the structure constants of the
group appear as momentum dependent phases. These phases cancel out in
planar diagrams. For non-planar diagrams,  the phases oscillate
very quickly for typical momentum values, since they are proportional to 
$\bar{k}L$, producing a strong suppression. The integer $\bar{k}$ is
defined by the relation  $k\bar{k} = 1 \mod L$. To diminish  
corrections from this source for low momentum modes, we proposed that
the large $N$ limit should be taken with both $k$ and $\bar{k}$ 
proportional to $L$.  A recent  analysis carried on for  the 2+1 
dimensional gauge theory~\cite{2+1,2+1P} shows that this 
prescription avoids also instabilities associated to the vacuum 
polarization (See Ref.~\cite{BNSV} for numerical results related to
this phenomenon).

Recently, space-time reduced models of large N QCD with adjoint fermions have 
attracted much attention\cite{KUY, AHUY, BS2, BKS, HN1, HN2}.  It is claimed that, in
the presence of adjoint fermions, the Z$^4$(N)  symmetry of the one-site model 
is not  broken even in the periodic boundary conditions case.  However,
for reduced  models with periodic  boundary conditions, there is not
such a clear connection  between the finite $N$ corrections and the finite volume
corrections of the ordinary lattice model. This makes it harder to determine  what
observables are most affected by finite N corrections, and what is the typical 
size of these corrections.

The purpose of the present letter is to construct the twisted space-time reduced model of 
large N QCD with two adjoint Wilson fermions and make a study of  its finite $N$ behavior.
The perturbative arguments relating the finite $N=L^2$ reduced model
with the lattice model on an  $L^4$ volume, extend to the case of
non-zero flavours of quarks in the adjoint representation. Our results
show that the $N$ dependence of various observables of the model with periodic 
boundary conditions ($k=0$) is large  and even irregular. On the contrary,
those obtained for our preferred values of $k$ show a mild dependence.  
In particular, for  $N=289=17^2$ and $k=3,5$, it is possible to calculate 
the Wilson loop $W(R,T)$ up to size $R,T=$8, and the results are  consistent 
with what is expected for  the large N theory  on a  $17^4$ lattice.

We start from the lattice action of the SU(N) gauge theory coupled
with  $N_f$ flavours of adjoint Wilson fermions,
given by
\begin{eqnarray}
%  \begin{split}
\nonumber
  S= & -b N \sum_n \sum_{\mu \ne \nu =1}^4 {\rm Tr} \left[  V_{n,\mu} V_{n+\mu,\nu}
 V_{n+\nu,\mu}^\dagger V_{n,\nu}^\dagger \right]     - 
    \sum_{j =1}^{N_f}  \sum_n {\rm Tr} \left[ \right. {\bar \Psi}_n^j \Psi_n^j
    -  \\
      &\kappa \sum_{\mu=1}^4  \{ {\bar \Psi}_n^j (1-\gamma_\mu) V_{n,\mu} 
      \Psi_{n+\mu}^j V_{n,\mu}^\dagger +{\bar \Psi}_n^j (1+\gamma_\mu) V_{n-\mu,\mu}^\dagger 
      \Psi_{n-\mu}^j V_{n-\mu,\mu}
      \} \left. \right]  
%  \end{split}  
\label{SLGT}
\end{eqnarray}
where $b$ is the inverse 't Hooft coupling on the lattice ($b=1/g^2N$). 
$V_{n,\mu}$ are SU(N) matrices, where $n$ runs over the points of a
4-dimensional lattice and $\mu$ over the directions. The fermionic fields
$\Psi_n^j$  are Grassman-valued $N\times N$ matrices, and transforming
by adjoint action of the SU(N) group. The label $j$ runs over the $N_f$=2 
flavour indices. In addition, these fields have spinor indices which
are not explicitly shown.  

In Ref.~\cite{TEK} we gave the prescription to construct the
twisted  reduced model for theories in which the different fields
transform in the adjoint representation of the group, which includes
the present model. The twisted reduced model can be obtained by 
applying the following replacement rules:
\begin{eqnarray}
%  \begin{split}
\nonumber
V_{n,\nu} &\rightarrow& V_\nu  \\
V_{n+\mu,\nu} &\rightarrow& \Gamma_\mu V_\nu \Gamma_\mu^\dagger
\nonumber \\
V_{n-\mu,\nu} &\rightarrow& \Gamma_\mu^\dagger V_\nu \Gamma_\mu
\nonumber\\
\Psi_n^j &\rightarrow& \Psi^j  \label{TR}\\
\Psi_{n+\mu}^j &\rightarrow& \Gamma_\mu \Psi^j \Gamma_\mu^\dagger
\nonumber\\
\Psi_{n-\mu}^j &\rightarrow& \Gamma_\mu^\dagger \Psi^j \Gamma_\mu .
\nonumber
%  \end{split}  
\end{eqnarray}
where the four twist matrices $\Gamma_\mu$ were first  introduced in 
ref.\cite{TEK} and satisfy the following algebra:
\begin{equation}
\Gamma_\nu \Gamma_\mu = z_{\mu\nu} \Gamma_\mu \Gamma_\nu .
\label{GAMMA}
\end{equation}
For the symmetric twist case, employed in this letter, we take $N=L^2$ with
$L$ a positive integer, and
\begin{equation}
 z_{\mu\nu} = \exp \left( k \frac{2\pi i}{ L} \right), \ \ \
 z_{\nu\mu}=z_{\mu\nu}^*, \ \ \ \mu>\nu
 \label{TT}
 \end{equation}
 The integer $k$ represents the flux through each plane. To preserve
 enough symmetry in the weak coupling  limit,  $k$ and $L$ should be
 co-prime. We recall that our prescription  to preserve the symmetry
  at intermediate couplings for the pure gauge  model~\cite{TEK2}, is 
  to take both  $k$ and $\bar{k}$ (defined earlier)  large enough.
 
After twisted reduction, there remain only four SU(N) matrices $V_\nu$ and
two adjoint fermionic matrix fields $\Psi^j$. Applying  the rules~(\ref{TR})
to the  action~(\ref{SLGT}), we obtain
\begin{eqnarray}
%  \begin{split}
\nonumber
   S=\ & -b N \sum_{\mu \ne \nu =1}^4 {\rm Tr} \left[ V_\mu \Gamma_\mu
  V_\nu \Gamma_\mu^\dagger \Gamma_\nu V_\mu^\dagger \Gamma_\nu^\dagger
  V_\nu^\dagger \right]
      - \sum_{j =1}^{N_f} {\rm Tr}\left[ \right. {\bar \Psi}^j \Psi^j \\
      \label{STR}
            -&\kappa \sum_{\mu=1}^4 \left\{ {\bar \Psi}^j
	    (1-\gamma_\mu) V_{\mu} \Gamma_\mu \Psi^j
	    \Gamma_\mu^\dagger V_{\mu}^\dagger
	     +{\bar \Psi}^j (1+\gamma_\mu) \Gamma_\mu^\dagger
	     V_{\mu}^\dagger \Psi^j V_{\mu} \Gamma_\mu
	         \right\} \left. \right].
%		   \end{split}
		   \end{eqnarray}
Finally, applying the change of variables  
\begin{equation}
U_\mu = V_\mu \Gamma_\mu
\label{UVG}
\end{equation}
and using (\ref{GAMMA}), we arrive at
\begin{eqnarray}
%  \begin{split}
\nonumber
   S=\ &- b N \sum_{\mu \ne \nu =1}^4 {\rm Tr} \left[ z_{\mu\nu} U_\mu
  U_\nu U_\mu^\dagger U_\nu^\dagger \right] 
      - \sum_{j =1}^{N_f} {\rm Tr}\left[ \right. {\bar \Psi}^j \Psi^j \\
            &-\kappa \sum_{\mu=1}^4 \left\{ {\bar \Psi}^j
	    (1-\gamma_\mu) U_{\mu} \Psi^j U_{\mu}^\dagger
	     +{\bar \Psi}^j (1+\gamma_\mu) U_{\mu}^\dagger \Psi^j U_{\mu}
	         \right\} \left. \right] \label{STR2} \\
		 \equiv\ &-b N \sum_{\mu \ne \nu =1}^4 {\rm Tr} \left[
		 z_{\mu\nu} U_\mu U_\nu U_\mu^\dagger U_\nu^\dagger \right]
		 - \sum_{j =1}^{N_f} {\rm Tr}\left[ {\bar \Psi}^j D_W
		 \Psi^j  \right] . \nonumber
%		   \end{split}
		     \end{eqnarray}
in which the  twist matrices $\Gamma_\mu$ have disappeared completely,
and there only remains a  $z_{\mu\nu}$ factor in the gauge part of the action.

The action~(\ref{STR2})  is the final form of the twisted reduced
model. Notice that for $k=0$ it becomes identical  to the reduced model with 
periodic boundary conditions ($\Gamma_\mu$=I for all $\mu$). The  action
is invariant under the Z$^4$(N) global symmetry $U_\mu \rightarrow z_\mu U_\mu$, 
with $z_\mu \in$Z(N). However, to preserve the equivalence of the reduced model 
with the original  lattice model, it is necessary to ensure that a
sufficiently large subgroup of the full group remains unbroken
spontaneously. In the pure gauge case the symmetric twist boundary
conditions are designed to preserve invariance under the subgroup Z$^4$(L)
in the weak-coupling limit. To monitor  that this subgroup  remains
unbroken, we always calculate the expectation value of  ${\rm Tr}
(U_\mu^\ell$), for $1\le l\le  (L-1)$,
which  are the  order parameters of the Z$^4$(L) symmetry. Indeed, we
confirm that, in all the simulations presented here, the quantities   $<{\rm Tr} (U_\mu^\ell)>$ are 
compatible with zero within statistical errors.  
%Although, not
%required for reduction to work, we also found that  $<{\rm Tr}
%(U_\mu^L)>=0$ is also compatible with zero. Thinking of this quantity
%as a Polyakov loop of an effective  lattice theory on a $L^4$
%lattice\cite{TEK}, this suggests that in all our simulations we remain  
%in the finite temperature confined phase.  
    
\begin{figure}[htb]
\begin{center}
\includegraphics[width=.6\textwidth]{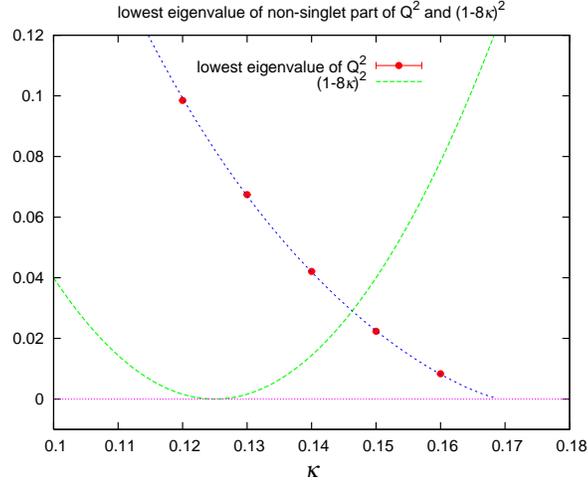}
\end{center}
\caption{$\kappa$ dependence of the lowest color non-singlet eigenvalue of $Q^2$ at $b$=0.35 and $k$=1.}
\label{fig1}
\end{figure}

In this work we have simulated the previous model for various values
of $b$, $\kappa$, $k$ and $N$ using the Hybrid Monte Carlo method.  
Introducing momenta $P_\mu$ conjugate to the link variables $U_\mu$ and the pseudofermion 
field $\phi$, the HMC Hamiltonian reads
\begin{equation}
 H=\frac{1}{2}  \sum_\mu {\rm Tr}(P_\mu^2) 
 - b N \sum_{\mu \ne \nu =1}^4 {\rm Tr} 
 \left(z_{\mu\nu}U_\mu U_\nu U_\mu^\dagger U_\nu^\dagger \right)
 + {\rm Tr}(\phi^\dagger Q^{-2}\phi) 
\label{H}
\end{equation}
where $Q=D_W \gamma_5=Q^\dagger$ is the hermitian Wilson Dirac matrix and $P_\mu$ are traceless $N \times N$
hermitian matrices.  $\phi$ is a complex $N \times N$ matrix with
implicit Dirac and colour indices (having $4N^2$  components).  $Q^2$ has $4N^2$ real and positive eigenvalues. 
Among them, four degenerate eigenvalues $(1-8\kappa)^2$ are related to
the unnecessary color singlet component  of the 
$N\otimes{\bar N}$  representation.   In fig. \ref{fig1}, we plot the lowest color non-singlet
eigenvalue  of $Q^2$ together with  $(1-8\kappa)^2$ at $b$=0.35 and $k$=1 for several values of $\kappa$.  It has been 
claimed\cite{BKS} that by imposing  ${\rm Tr}\phi$=0, we can reduce the number of CG iterations.  
This is true if $(1-8\kappa)^2$ is smaller than the lowest non-singlet eigenvalues of $Q^2$.  Otherwise, 
the color singlet part of $N\otimes{\bar N}$ has practically no effect at all.  Since our ultimate goal
is to study the small quark mass region at large values of $\kappa$, we do not impose the
constraint ${\rm Tr}\phi$=0 in our simulations.

We invert $Q^2$ using the conjugate gradient (CG) algorithm.  Our stopping condition is as follows.
Let $r = s-Q^2 x$ be the residue with $s$ the source.  Then we require $|r|^2/|s|^2 < 10^{-7}$ during 
the molecular dynamics evolutions, and  $|r|^2/|s|^2 < 10^{-15}$ at the global reject-accept step.
We always start from $x_0$=0 for the initial value of a solution vector to guarantee the reversibility of 
the trajectories.  We use trajectories with unit length, so that the  step size $\Delta \tau$ in the molecular
dynamics evolution and the number of  time steps  NMD within one
trajectory are  inversely connected, $\Delta \tau =$1/NMD. We measure observables every 10 trajectories.
     
Simulations have been done on Hitachi SR16000 super computer with one node 
having 32-cores IBM power7 and peak speed of 980 GFlops.  Our codes are highly optimized for SR16000 and
the sustained speed is 300-600 GFlops depending on the values of $N$
and $\kappa$.   The use of this high speed  supercomputer enables us to simulate the model up to $N$=289, 
whereas  previous simulations for $k=0$ went  only up to $N<50$\cite{AHUY,BKS}.   

\begin{figure}[htb]
\begin{center}
\includegraphics[width=.6\textwidth]{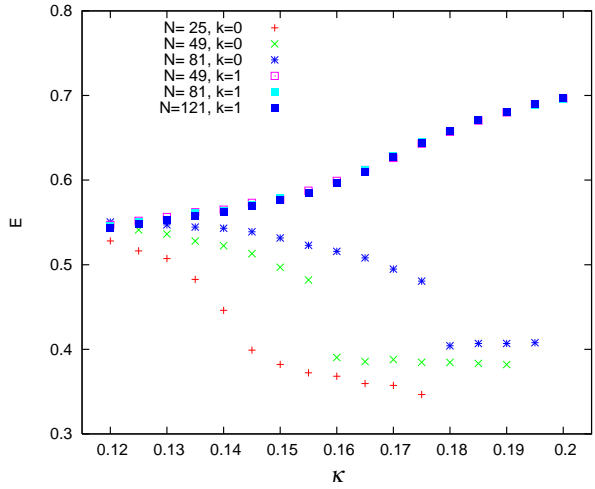}
\end{center}
\caption{$\kappa$ dependence of the plaquette expectation value $E$
for several values of $N$ 
at $b$=0.35 and $k$=0 and 1.}
\label{fig2}
\end{figure}

Our first analysis focuses on the expectation value of the plaquette 
\newline
$E=<z_{\mu\nu}\mathrm{Tr}(U_\mu U_\nu U_\mu^\dagger U_\nu^\dagger)>$ and its dependence
on $\kappa$ and $N$.  We first made a rapid scan of the system at $b$=0.35.  
In fig. \ref{fig2}, we plot the $\kappa$ dependence of the  
plaquette value $E$  both for $k$=0 and 1 with  various values of $N$.  For each $N$ and $k$, 
we start from $\kappa$=0.12 and increase $\kappa$ in steps of  0.005.  
At each $\kappa$, we run 500 trajectories and make  averages using the last 300 trajectories.

For the periodic boundary conditions case $k$=0, the results
of our simulations performed  at  $N$=25, 49 and 81, show that 
as $\kappa$ increases $E$ decreases, which is quite unnatural.  One usually expects to have 
larger dynamical quark effects as $\kappa$ increases (quark mass decreases).  Since the dynamical
quark effects tend to order the system, we usually expect to have larger plaquette value $E$ 
with larger $\kappa$.  Notice also that there exists an abrupt change of $E$
at an   intermediate value of $\kappa$.  The value at which this jump
takes place increases by a sizable amount as  $N$ grows. Hence, it is
clear that this feature should be a finite $N$ artifact. The reason
for this behaviour is unclear, but we know that it is  not associated 
to  any  breaking of the Z$^4$(L) symmetry since, as stated earlier, 
there is no evidence  of breaking  in all the simulations presented 
in this paper. 

In contrast the $E$ dependence of the twisted $k=1$ model looks very
different. In this case the plaquette expectation value $E$ grows with
$\kappa$ as expected. Furthermore, the $N$ dependence is much smaller
and difficult to see in this rough scale. Notice also that the
$k=0$ curves seem to approach the $k=1$ result as $N$ increases. Thus, 
it is a priori possible  that for asymptotic values of $N$ all the curves 
match.

To understand if this is indeed the case, we made a precise study of
the $N$ dependence of the plaquette expectation value $E$ at three
fixed values of the parameters ($b$ and $\kappa$) of the action. We
chose  ($b$, $\kappa$)=(0.35, 0.12), (0.36, 0.12) and (0.35, 0.14). The 
corresponding results are compiled in tables~\ref{table1}-\ref{table3}. 
For each run, we use 4000 trajectories, discarding at least the initial 
1000 trajectories for thermalization.  Errors are estimated by the
jack-knife method, splitting up the  4000 trajectories in 10 bins.  We choose NMD so as to make the 
global acceptance ratio larger than 0.5, except for two runs at $b$=0.35 and 
$\kappa$=0.14.  After we started these simulations, we realized that the low acceptance
ratio is not due to  the small value of NMD, but rather due to the weaker stopping 
condition $|r|^2/|s|^2 < 10^{-7}$.  
Although a low acceptance ratio does not introduce any systematic error 
since HMC algorithm is exact, it is desirable to have configuration sets with 
short autocorrelations. Hence, in the future simulations, needed for 
the computation of the string tension, we will be able to get higher 
values of the  acceptance ratio.

\begin{figure*}[htb]
\centering
\subfigure{\includegraphics[width=.45\textwidth]{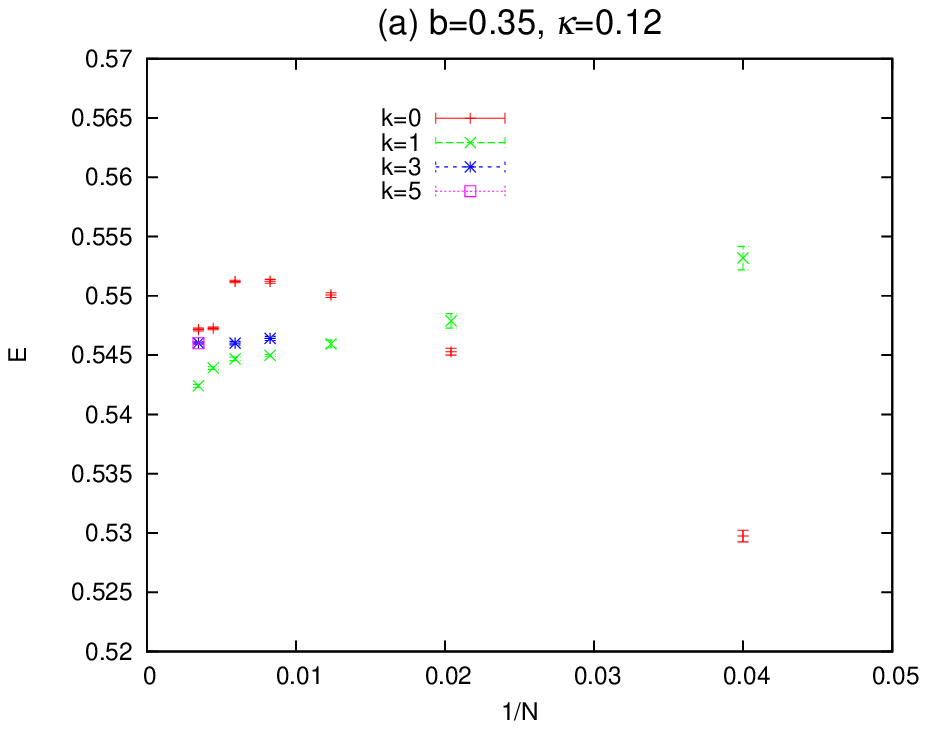}}
\vspace{-1.5cm}
\\
\subfigure{\includegraphics[width=.45\textwidth]{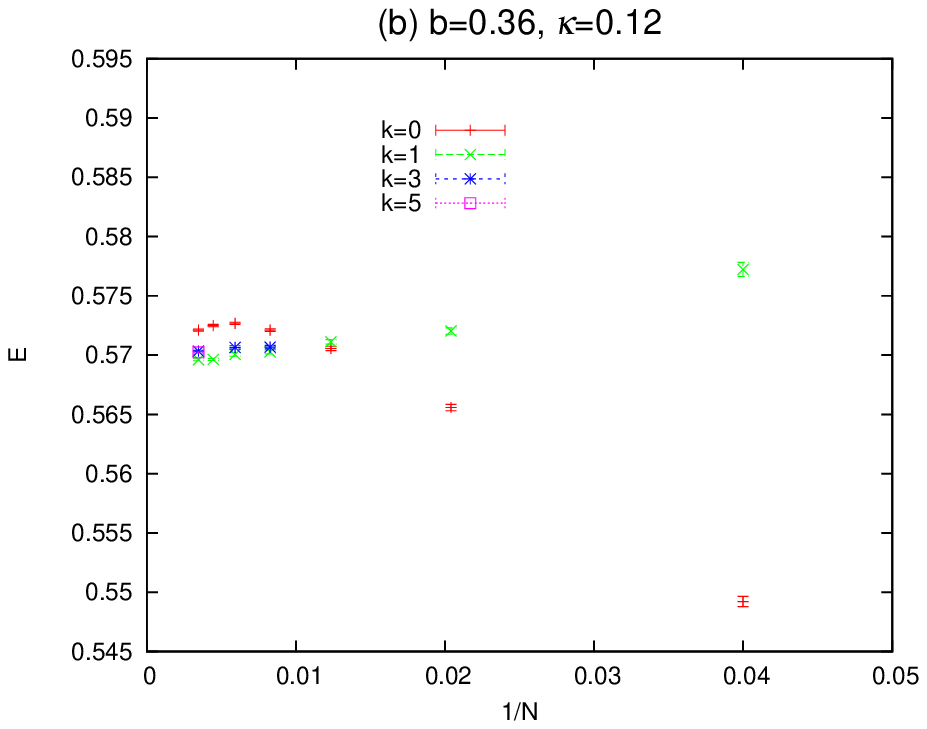}}
\vspace{-1.5cm}
\\
\subfigure{\includegraphics[width=.45\textwidth]{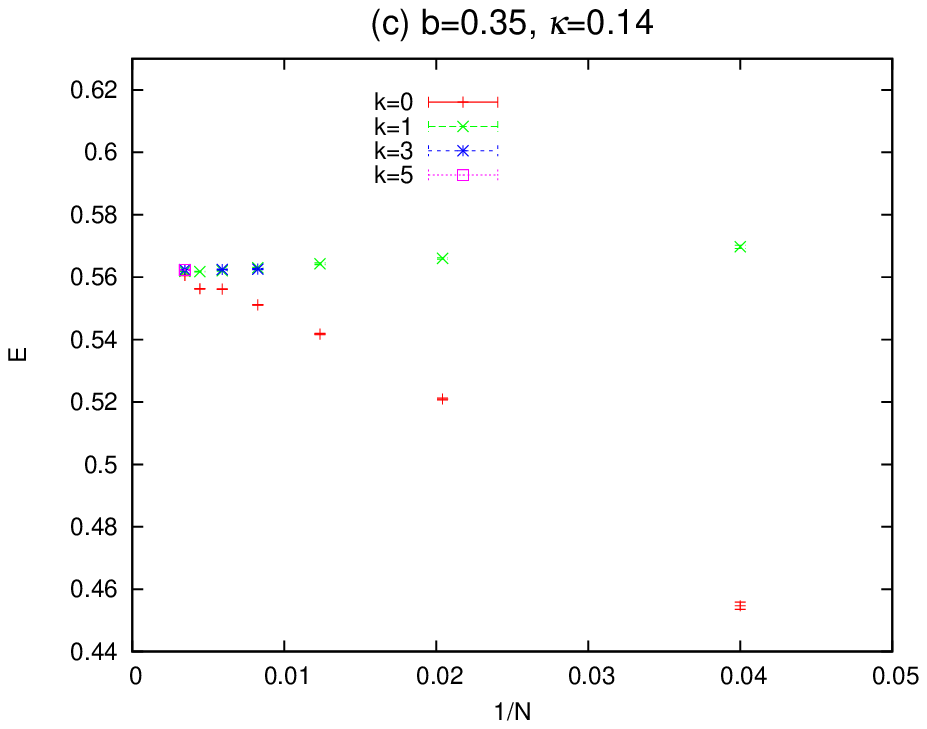}}
\vspace{-1.cm}
\caption{$1/N$ dependence of $E$ for various $k$.}
\label{fig3}
\end{figure*}

Let us now examine our results. We begin by those obtained at
$b$=0.35 and $\kappa$=0.12, presented in table~\ref{table1} and
displayed in fig.~\ref{fig3}(a). The results obtained for large $N$ 
and $k=3$ and $k=5$ are consistent with each other and give a value 
of $E=$0.5460(2). On the other hand, the results obtained for $k=0$
(periodic boundary conditions) and $k=1$ show an irregular behaviour. 
For the  $k=0$ case, there is first a growth and then a decrease as a
function of $N$. Given that $\kappa$ is small, this might be related 
to the instability of the vacuum for the pure gauge theory. For $k=1$
the instability of the pure gauge theory sets in at
$N>100$~\cite{IO, TV, AHHI} and this
might also explain the apparent change of tendency present  in the data.
Indeed, the two smallest values of $N$ for the $k=1$ fall into the line 
$0.5460+4.47/N^2$. However, for larger values of $N$ the value of $E$
does not show signs of stabilizing close to our preferred result. For
$k=3$, symmetry breaking for the pure gauge theory sets in at $N>784$,
and, if our conjecture is correct,  there could be a similar behaviour
to be expected at those values  of $N$, which are beyond the reach of our
computational power. For $k=5$  no breaking has been seen in the pure
gauge theory, but according to our arguments of Ref.~\cite{TEK2} it should
appear for $N>2200$. 

In any case, our conclusion is that it is wise to use the same criteria 
for the model with adjoint fermions as for the pure gauge case since
it comes with no additional cost. Namely one has to take $k$ and $L$
coprime, $k/L >0.1$ and with ${\bar k}/L$ large enough. Our $k=$3 and 5
results satisfy these criteria. 

\begin{figure}[htb]
\begin{center}
\includegraphics[width=.6\textwidth]{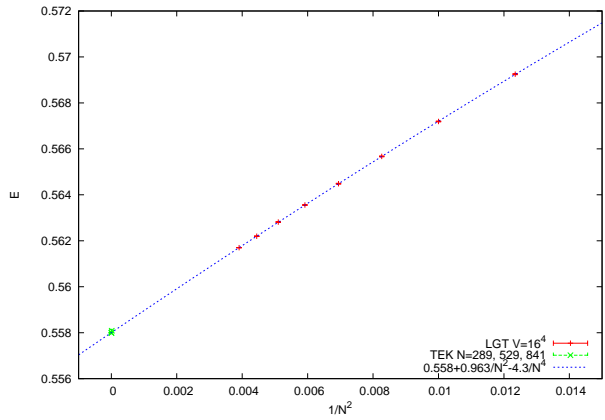}
\end{center}
\caption{Plaquette expectation value $E$ for the pure gauge case as a
function of $1/N^2$. The data for $N\le 16$ are obtained for 
ordinary lattice gauge theory on a  $16^4$ lattice. The continuous line 
shows the result of a fit, which allows the extrapolation to infinite
N. On the same plot we also display the value of $E$ obtained for 
the TEK model at N=841, 529 and 289.}
\label{figTEK}
\end{figure}

It is interesting to explore what happens at the other values of $b$ and
$\kappa$. The results of $(b,\kappa)=(0.36,0.12)$ are collected in
table~\ref{table2}  and fig.~\ref{fig3}(b), and those of 
$(b,\kappa)=(0.35,0.14)$ in table~\ref{table3}  and fig.~\ref{fig3}(c).
At these larger values of $b$ and $\kappa$ things seem to improve for
the $k=1$ twisted case and get somewhat worse for the $k=0$ one. 
For example, if we apply the same criteria to the data at $b=0.36$ as before, the
correct result for the plaquette expectation value would be
$E=0.5702(1)$. This is essentially the result for $k=5$, while the
data of $k=3$ can be fitted to $E=0.5702(1)+9(2)/N^2$. If we fit the 
$k=1$ data for $N<200$ we get $0.5700(2)+4.17/N^2$  with a chi square 
per degree of freedom equal to 1. Indeed, the two largest values of $N$
give a slightly lower value of $0.5696(1)$, probably due to the same
problem observed before. On the contrary the $k=0$ has a very strange
behaviour. For small $N$, the finite $N$ corrections  are huge
compared to the twisted case, but then seems to settle at a value 
of 0.5725(3), although there might be hints that it is starting to
decrease. The small difference of $0.002$ in the plaquette value for 
periodic and twisted boundary conditions might seem worrisome. Our 
confidence  in the level of precision of the twisted result is based on the
analysis of the pure gauge case. It is then  possible to compare the plaquette
expectation value obtained for the twisted Eguchi-Kawai model at b=0.36 
with the result of extrapolating the plaquette expectation value for
ordinary lattice gauge theory on a $16^4$ box (with periodic boundary
conditions) and different values of $N$. Indeed, we obtained data for 
all values of $N$ in the range 9-16 (8 values) with sufficient
statistics to achieve a precision of around 0.00002 in all cases. Then
we fitted the results to a second degree polynomial in $1/N^2$ (3
parameters). The fit is very good with a chi square per degree of
freedom of order  1. The best fit gives
$E=0.55800(2)+0.963(6)/N^2-4.3(4)/N^4$. The extrapolated result matches 
the value obtained for the TEK model for $N=841$ and $k=9$, given by 0.557998(5). 
Compatible results with larger errors were obtained for $N=$289 with
$k=5$ ($E=$0.557991(18)), and 529 with $k=7$ ($E=$0.557989(13)).
All the mentioned data and the best fit function are displayed in
fig.~\ref{figTEK}. Statistical errors, also displayed, are too small
to be seen on the scale of the figure. 

Of course, all the previous information affects the pure gauge model,
but there is no  reason why the presence of fermions in the
adjoint representation should spoil this agreement, but just the
opposite.

The data at $b=0.35$, $\kappa=0.14$ confirms the picture. The $k=5$ 
value is 0.56235(1), the $1/N^2$ fit to $k=3$ gives 0.56234(2), 
and a 3 parameter  fit (second degree polynomial in $1/N^2$) to the
4 smallest N values of $k=1$ gives 0.5626(3) as large N predictions 
for the plaquette value. Again the larger values of $N$ for $k=1$ 
give slightly lower values for the plaquette. The data of $k=0$
shows a very strong and irregular N-dependence, with  a similar growth 
for small N, followed by apparent  stabilization and a second growth.

Let us now explore the behaviour of larger Wilson loops. A priori 
one expects that the finite $N$ corrections grow with the loop size. 
For the twisted model, as mentioned previously, perturbation theory
(and other considerations) relate the finite $N=L^2$ theory with the
lattice gauge theory at finite volume $L^4$.   Hence, we expect  the 
corrections to become fairly large when the size of the loop is of the 
order of $L/2$.  If the size is smaller one could obtain reasonably 
good approximations  to the expectation values. For the periodic 
boundary conditions case  $k=0$ it is unclear how these affect the 
observables with a characteristic lattice scale. In what follows we will 
present our results which should explore these matters. 

\begin{figure*}[htb]
\centering
\subfigure{\includegraphics[width=.45\textwidth]{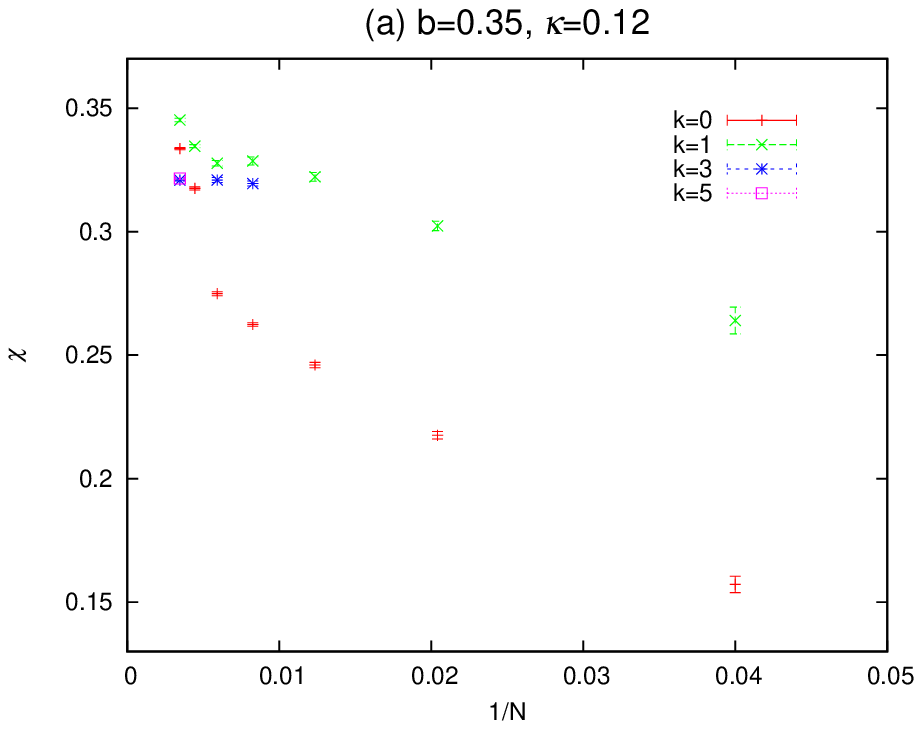}}
\vspace{-1.5cm}
\\
\subfigure{\includegraphics[width=.45\textwidth]{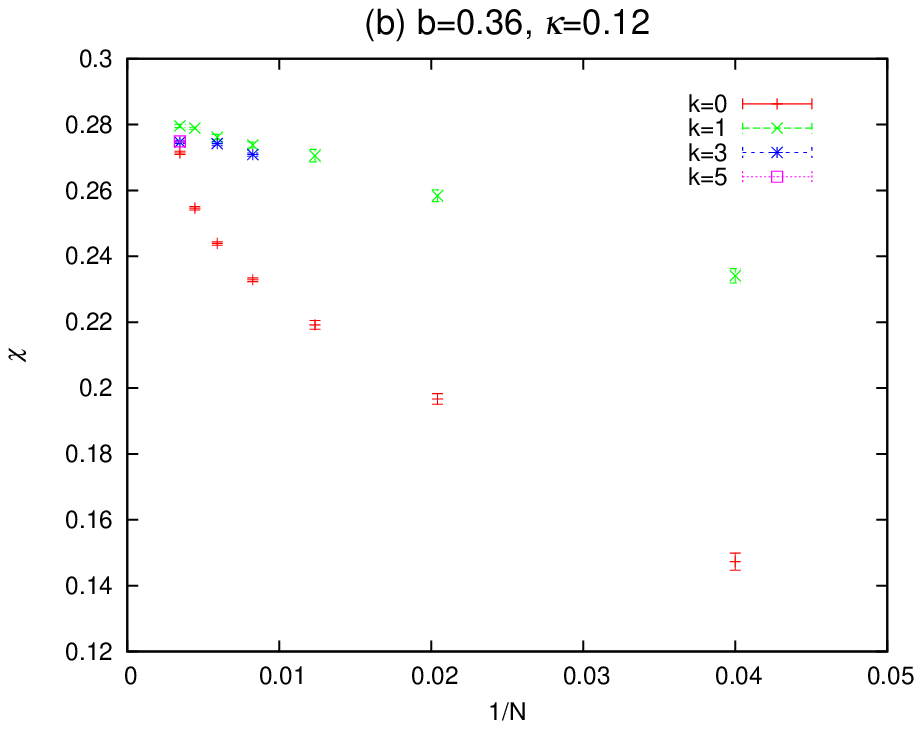}}
\vspace{-1.5cm}
\\
\subfigure{\includegraphics[width=.45\textwidth]{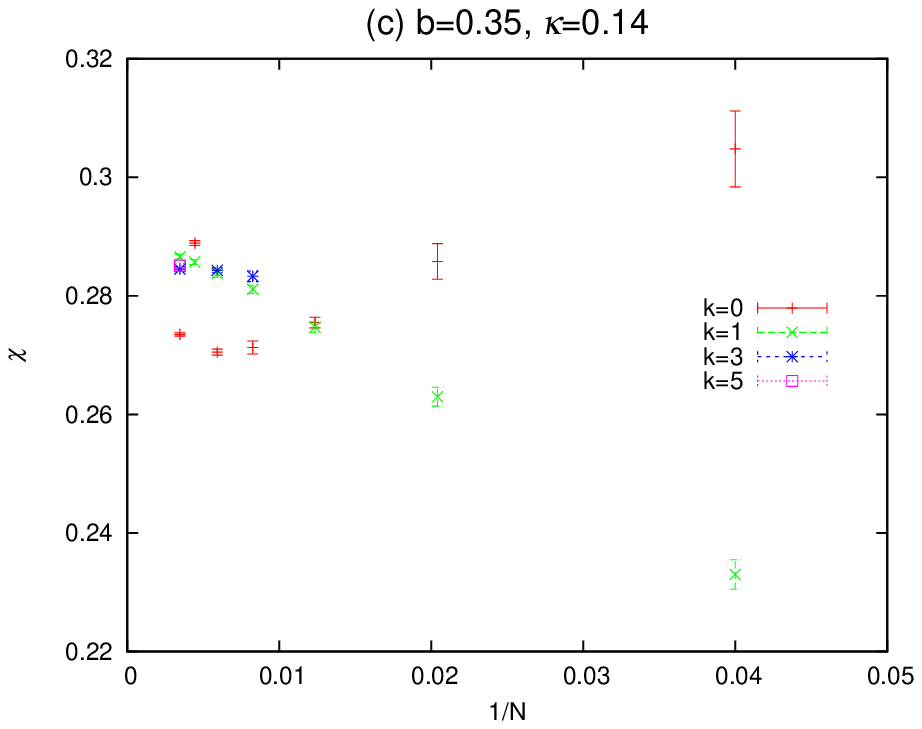}}
\vspace{-1.cm}
\caption{$1/N$ dependence of $\chi$ for various $k$.}
\label{fig4}
\end{figure*}

In the twisted reduced model, the Wilson loop $W(R,T)$ with size 
$R \times T$ is calculated from
the four link variables $U_\mu$ as 
\begin{equation}
 W(R,T) =  z_{\mu\nu}^{RT} < \mathrm{Tr}(U_\mu^R U_\nu^T U_\mu^{R
 \dagger} U_\nu^{T \dagger}) >.
\label{WILSON} 
\end{equation}
However,  Wilson loops are divergent quantities in the continuum
limit. One can eliminate constant and perimeter divergences by forming 
Creutz ratios defined as 
\begin{equation}
\chi(R+1/2,T+1/2) = -\log{ \frac{W(R+1,T+1) W(R,T)}{ W(R+1,T) W(R,T+1)} }.
\label{CHI}
\end{equation}
Creutz ratios are therefore very important observables, from which one
can extract continuum limit quantities, such as the string tension and
other continuum functions (See Ref.~\cite{TEK3,TEK4}). Thus, we will start 
by presenting the $N$ dependence observed in our study for the smallest ratio 
$\chi(3/2,3/2)$, focusing on the three $(b,\kappa)$ used
before for the plaquette. The numerical values appear in the last
column of the same tables (\ref{table1}-\ref{table3}) used before. 
The data is also plotted in fig. \ref{fig4} to allow for a visual
comparison. 

The results are in line with our previous findings for the plaquette. 
The data for $k=3$ and $k=5$ and the various $N$ agree among
themselves. For the other values of $k$ the initial trend is towards 
approaching, but at higher values of $N$ differences are clearly
present. The behaviour becomes particularly irregular for $k=0$ and 
$\kappa=0.14$ for which there is a sudden jump at $N$=225. This
strange result might suggest that there are different phases and the
system might flip from one to other at different $N$ values.

\begin{figure}[htb]
\begin{center}
\includegraphics[width=.6\textwidth]{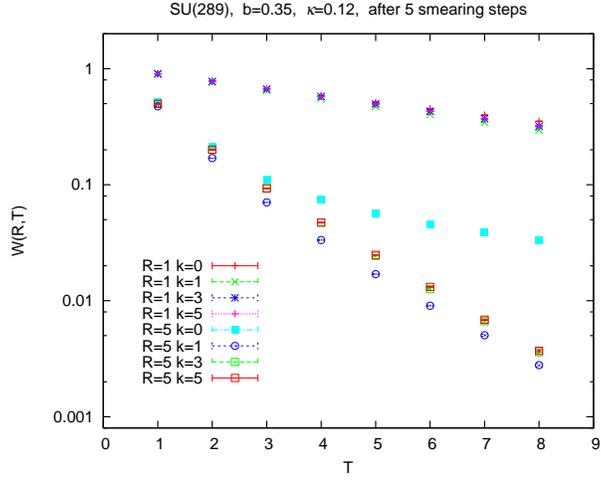}
\end{center}
\caption{Wilson loop for various values of $k$ after 5 smearing steps.}
\label{fig5}
\end{figure}

To study even larger size Wilson loops one has to face the problem that 
the errors tend to increase considerably. To reduce the fluctuations
we will make use of the well-known  smearing technique.
Here we use the conventional Ape smearing\cite{APE}
\begin{equation}
 U_\mu^{smeared}={\rm Proj}_N \left[ U_\mu + c \sum_{\nu \ne \mu}(z_{\nu \mu} U_\nu U_\mu U_\nu^\dagger 
                + z_{\mu \nu} U_\nu^\dagger U_\mu U_\nu) \right]
\label{SM}  
\end{equation}    
where ${\rm Proj}_N$ stands for the projection  operator  onto  SU(N) matrices.    
In fig. \ref{fig5}, we show the $k$ dependence of the Wilson loops at
$N$=289, $k$=5, $b$=0.35 and $\kappa$=0.12 after 5 smearing steps with the smearing parameter 
$c$=0.1. For $R$=1, the Wilson loops  $W(R=1,T)$  seem to  coincide
for all $k$ (0, 1, 3 and 5) within  such rough scale.  For $R$=5, the  Wilson loops $W(R=5,T)$ 
calculated with $k$=3 and 5 agree with each other.  According to our
previous  arguments,  they are indeed   the correct values  of the large N theory.  
It is amazing that  the Wilson loops calculated with  the periodic boundary condition ($k$=0) 
for $T\ge 4$ are so  different from those with $k$= 3  and 5 even at $N$=289.  The results 
with $k$=1 are also smaller than those of  $k$= 3 and 5.

\begin{figure}[htb]
\begin{center}
\includegraphics[width=.6\textwidth]{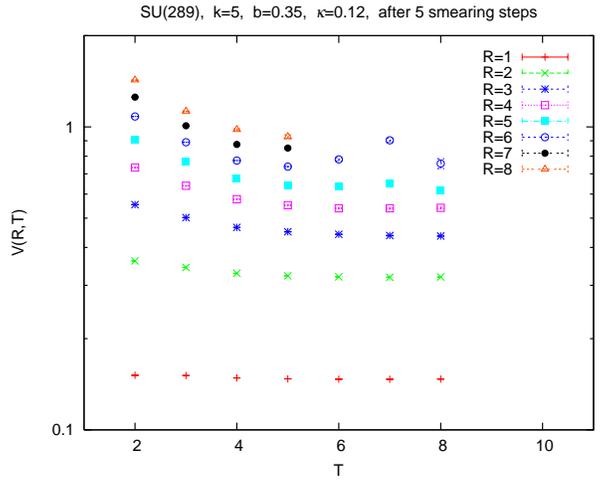}
\end{center}
\caption{Quark potential $V(R,T)$ after 5 smearing steps.}
\label{fig6}
\end{figure}

Finally, we discuss the quark potential defined by
\begin{equation}
V(R,T) = -\log{ \frac{W(R,T)}{ W(R,T-1)} },
\label{VRT}
\end{equation}     
which is also important from the phenomenological point of view.  We
expect that for large $T$, $V(R,T)$ tends to  $V(R)$, representing 
the $q {\bar q}$ potential at separation $R$. In fig. \ref{fig6}, we show the
result of $V(R,T)$ at $N$=289, $k$=5, $b$=0.35, $\kappa$=0.12 after 5
smearing steps.   We clearly observe a plateau for large $T$ up to  $T$=8 with $R=1 \sim 6$.   
For $R$=7 and 8, we cannot calculate $V(R,T)$ for $T\ge 6$ due to the large noises, 
but this  problem is not specific of the reduced model, since it also
appears  in ordinary lattice gauge theory. Thus, our results show that, indeed,
the reduced model behaves as the large N lattice theory with lattice size $17^4$.   
  
In conclusion, in this work we have presented and analyzed  the 
twisted reduced model of  large N QCD with two adjoint Wilson fermions. 
In particular, we focused in studying the $k$ and $N$ dependence of various 
quantities. Our results show that it is advisable to apply the same criteria
in selecting  the flux value $k$ as advocated for the pure gauge theory.
All the results obtained for that case are perfectly stable with $N$
and behave as expected. For instance, the results for $N$=289 and
$k=5$ show that indeed one can compute  Wilson loops $W(R,T)$ up to
$R,T$=8, in accordance with the association of finite $N$ effects with
finite volume effects on a $17^4$ lattice. The capacity to explore
large loops enables one to estimate the string tension for this theory. 
Some preliminary results obtained by us have already been
reported~\cite{MO}.

The situation for other values of $k$, including periodic boundary
conditions, is unclear. Although no symmetry breaking is observed for
the range of parameter values explored in this paper, they show large
N dependences and anomalous or irregular behaviour for large $N$. The
situation is particularly severe for the $k=0$ case. At present, the
interpretation is unclear. A recent work~\cite{LN} suggests that, at
least for  large values of $b$, the $k=0$ reduced model might fail to 
match the large N infinite volume theory except for strictly zero mass. 
It is interesting to point that these problems do not apply for the 
twisted version with $k\ne 0$. On the contrary, the large $b$ analysis 
shows that the distribution of eigenvalues is flat, as it should.
Our work shows that,  even if large N  reduction of the $k=0$ model 
indeed takes place for the region of  parameters analyzed in this work,
the large N dependences and irregularities presented here seriously 
limit the applicability  of the untwisted version of the reduced model.

\vspace{0.5cm}
\noindent
{\bf Acknowledgements}

The authors benefited from interesting conversations with Margarita Garcia Perez,  Ken-Ichi Ishikawa, 
Liam Keegan and Mateusz Koren.
    
A.G-A is supported from Spanish grants FPA2009-08785, FPA2009-09017, CSD2007-00042, HEPHACOS S2009/ESP-1473,
PITN-GA-2009-238353 (ITN STRONGnet), CPAN CSD2007-00042. and  the Spanish MINECO Centro de excelencia Severo Ochoa
Program under grant SEV-2012-0249. M.O is supported in part by Grants-in-Aid for Scientific Research from the 
Ministry  of Education, Culture, Sports, Science and Technology (No 23540310).

The calculation has been done on Hitachi SR16000-M1 computer at High Energy Accelerator
Research Organization (KEK) supported by the Large Scale Simulation Program No.12/13-01
(FY2012-13).  The authors thank the Hitachi system engineers for their help in highly optimizing the present simulation code.

\newpage

\begin{table}[htb]
\begin{center}
\begin{tabular}{|r|r|c|c|c|c|l|l|} \hline
  $k$  & $N$ & NMD & acceptance & NCGH & NCGF & \ \ \ \ \ \ $E$ & \ \ \ \ \ \ $\chi$ \\ \hline \hline
       &  25 & 200 & 0.966 & 77 & 44.7 & 0.52954(73) & 0.1572(33) \\ \cline{2-8}
       &  49 & 200 & 0.919 & 77 & 43.7 & 0.54529(29) & 0.2176(15) \\ \cline{2-8}
       &  81 & 200 & 0.849 & 75 & 42.5 & 0.54966(18) & 0.2460(11) \\ \cline{2-8}
     0 & 121 & 200 & 0.725 & 74 & 41.6 & 0.55117(22) & 0.2624(7)  \\ \cline{2-8} 
       & 169 & 200 & 0.516 & 73 & 40.3 & 0.55122(10) & 0.2749(8)  \\ \cline{2-8}  
       & 225 & 400 & 0.649 & 71 & 36.1 & 0.54726(11) & 0.3175(6)  \\ \cline{2-8}   
       & 289 & 400 & 0.544 & 71 & 29.9 & 0.54716(12) & 0.3336(4)  \\ \hline \hline
       &  25 & 200 & 0.982 & 54 & 32.6 & 0.55319(99) & 0.2640(54) \\ \cline{2-8}
       &  49 & 200 & 0.937 & 58 & 34.3 & 0.54789(61) & 0.3023(19) \\ \cline{2-8}
       &  81 & 200 & 0.871 & 60 & 34.9 & 0.54559(46) & 0.3222(18) \\ \cline{2-8}
     1 & 121 & 200 & 0.744 & 60 & 34.9 & 0.54509(23) & 0.3286(16) \\ \cline{2-8} 
       & 169 & 200 & 0.542 & 61 & 34.0 & 0.54469(15) & 0.3277(11) \\ \cline{2-8}  
       & 225 & 400 & 0.695 & 62 & 30.7 & 0.54392(11) & 0.3346(6)  \\ \cline{2-8}   
       & 289 & 400 & 0.562 & 63 & 25.3 & 0.54242(12) & 0.3452(7)  \\ \hline \hline
       & 121 & 200 & 0.742 & 60 & 34.8 & 0.54612(18) & 0.3195(10) \\ \cline{2-8}  
     3 & 169 & 200 & 0.557 & 60 & 33.6 & 0.54602(15) & 0.3209(10) \\ \cline{2-8}   
       & 289 & 400 & 0.567 & 60 & 24.0 & 0.54603(10) & 0.3209(4)  \\ \hline \hline
     5 & 289 & 600 & 0.638 & 60 & 24.0 & 0.54601(10) & 0.3215(4)  \\ \hline         
  \end{tabular}
\end{center} 
 \caption{We show several run parameters and observables for the
 simulations at $b$=0.35 and $\kappa$=0.12. The last two columns 
 list the value of  the plaquette $E$ and the Creutz ratio $\chi\equiv
 \chi(3/2,3/2)$. NMD is the number of time steps within one trajectory and 
 {\em acceptance} stands for the  global acceptance ratio at reject-accept
 step.  NCGH is the number of CG iterations at global reject-accept step 
 and NCGF is the number of CG iterations during the molecular dynamics evolutions.}
\label{table1}
\end{table}

\begin{table}[htb]
\begin{center}
\begin{tabular}{|r|r|c|c|c|c|l|l|} \hline
  $k$  & $N$ & NMD & acceptance & NCGH & NCGF & \ \ \ \ \ \ $E$ & \ \ \ \ \ \ $\chi$ \\ \hline \hline
       &  25 & 200 & 0.965 & 80 & 46.3 & 0.54842(63) & 0.1473(26) \\ \cline{2-8}
       &  49 & 200 & 0.917 & 80 & 45.3 & 0.56536(30) & 0.1967(16) \\ \cline{2-8}
       &  81 & 200 & 0.839 & 79 & 44.0 & 0.57046(28) & 0.2192(13) \\ \cline{2-8}
     0 & 121 & 200 & 0.720 & 77 & 43.2 & 0.57229(13) & 0.2329(6)  \\ \cline{2-8} 
       & 169 & 200 & 0.529 & 76 & 42.0 & 0.57268(10) & 0.2439(5)  \\ \cline{2-8}  
       & 225 & 400 & 0.641 & 75 & 38.4 & 0.57250(10) & 0.2546(5)  \\ \cline{2-8}   
       & 289 & 400 & 0.520 & 75 & 32.2 & 0.57211(9)  & 0.2714(5)  \\ \hline \hline
       &  25 & 200 & 0.975 & 55 & 33.0 & 0.57666(43) & 0.2341(22) \\ \cline{2-8}
       &  49 & 200 & 0.940 & 60 & 35.1 & 0.57184(35) & 0.2584(18) \\ \cline{2-8}
       &  81 & 200 & 0.862 & 62 & 36.0 & 0.57110(29) & 0.2706(19) \\ \cline{2-8}
     1 & 121 & 200 & 0.737 & 63 & 36.2 & 0.57040(15) & 0.2738(9)  \\ \cline{2-8} 
       & 169 & 200 & 0.540 & 63 & 35.5 & 0.57006(15) & 0.2762(8)  \\ \cline{2-8}  
       & 225 & 400 & 0.671 & 64 & 32.1 & 0.56964(9)  & 0.2789(3)  \\ \cline{2-8}   
       & 289 & 400 & 0.541 & 65 & 26.7 & 0.56961(9)  & 0.2796(5)  \\ \hline \hline
       & 121 & 200 & 0.731 & 62 & 35.9 & 0.57084(19) & 0.2709(5)  \\ \cline{2-8}  
     3 & 169 & 200 & 0.521 & 63 & 34.8 & 0.57065(15) & 0.2742(6)  \\ \cline{2-8}   
       & 289 & 400 & 0.535 & 63 & 25.4 & 0.57035(7)  & 0.2747(4)  \\ \hline \hline
     5 & 289 & 600 & 0.635 & 63 & 25.4 & 0.57024(5)  & 0.2749(3)  \\ \hline         
  \end{tabular}
\end{center} 
 \caption{The  parameters of the run, the average  value of the plaquette $E$ 
 and the Creutz ratio $\chi$ for the simulations at  $b$=0.36 and $\kappa$=0.12.
The meaning of the headings  is the same as in Table \ref{table1}.}
\label{table2}
\end{table}

\begin{table}[htb]
\begin{center}
\begin{tabular}{|r|r|c|c|c|c|l|l|} \hline
  $k$  & $N$ & NMD & acceptance & NCGH & NCGF & \ \ \ \ \ \ $E$ & \ \ \ \ \ \ $\chi$ \\ \hline \hline
       &  25 & 100 & 0.928 &  96 & 45.6 & 0.45516(153) & 0.3048(64) \\ \cline{2-8}
       &  49 & 100 & 0.763 & 110 & 46.4 & 0.52043(43)  & 0.2858(30) \\ \cline{2-8}
       &  81 & 200 & 0.798 & 112 & 44.4 & 0.54153(38)  & 0.2755(9)  \\ \cline{2-8}
     0 & 121 & 300 & 0.746 & 111 & 43.1 & 0.55135(5)   & 0.2713(11) \\ \cline{2-8} 
       & 169 & 300 & 0.613 & 110 & 41.7 & 0.55615(10)  & 0.2705(5)  \\ \cline{2-8}  
       & 225 & 400 & 0.532 & 108 & 40.1 & 0.55627(8)   & 0.2889(4)  \\ \cline{2-8}   
       & 289 & 500 & 0.427 & 108 & 39.5 & 0.56059(6)   & 0.2735(3)  \\ \hline \hline
       &  25 & 100 & 0.939 &  68 & 35.9 & 0.56960(82)  & 0.2330(25) \\ \cline{2-8}
       &  49 & 100 & 0.796 &  76 & 38.4 & 0.56600(30)  & 0.2630(16) \\ \cline{2-8}
       &  81 & 100 & 0.512 &  80 & 39.4 & 0.56420(26)  & 0.2747(9)  \\ \cline{2-8}
     1 & 121 & 200 & 0.667 &  82 & 39.6 & 0.56315(29)  & 0.2811(7)  \\ \cline{2-8} 
       & 169 & 300 & 0.627 &  83 & 38.9 & 0.56206(14)  & 0.2838(7)  \\ \cline{2-8}  
       & 225 & 400 & 0.549 &  84 & 37.2 & 0.56181(11)  & 0.2857(4)  \\ \cline{2-8}   
       & 289 & 500 & 0.468 &  86 & 34.9 & 0.56184(11)  & 0.2866(4)  \\ \hline \hline
       & 121 & 200 & 0.666 &  81 & 39.2 & 0.56260(19)  & 0.2833(9)  \\ \cline{2-8}  
     3 & 169 & 300 & 0.629 &  82 & 38.6 & 0.56244(19)  & 0.2843(4)  \\ \cline{2-8}   
       & 289 & 500 & 0.454 &  82 & 34.8 & 0.56239(11)  & 0.2845(4)  \\ \hline \hline
     5 & 289 & 800 & 0.500 &  82 & 34.8 & 0.56235(7)   & 0.2851(4)  \\ \hline         
  \end{tabular}
\end{center} 
 \caption{Run parameters,  plaquette value $E$ and  Creutz ratio $\chi$ 
at $b$=0.35 and $\kappa$=0.14.
The meaning of run parameters is the same as in Table \ref{table1}.}
\label{table3}
\end{table}

%\begin{figure*}[htb]
%\centering
%\subfigure{\includegraphics[width=.45\textwidth]{fig6_1.eps}}
%\vspace{-1.5cm}
%\\
%\subfigure{\includegraphics[width=.45\textwidth]{fig6_2.eps}}
%\vspace{-1.5cm}
%\\
%\subfigure{\includegraphics[width=.45\textwidth]{fig6_3.eps}}
%\vspace{-1.cm}
%\caption{$1/N^2$ dependence of $E$ at $k$= 3 and 5 for various $b$ and $\kappa$.}
%\label{fig6}
%\end{figure*} 

\vskip 0.5cm

\end{document}